\documentclass[conference]{IEEEtran}
\IEEEoverridecommandlockouts
\usepackage{cite}
\usepackage{amsmath,amssymb,amsfonts}
\usepackage{algorithmic}
\usepackage{graphicx}
\usepackage{textcomp}
\usepackage{xcolor}
\usepackage{url}
\usepackage{tikz}
\usetikzlibrary{shapes,arrows,positioning,fit,backgrounds,calc,shadows}
\usepackage{caption}
\def\BibTeX{{\rm B\kern-.05em{\sc i\kern-.025em b}\kern-.08em
    T\kern-.1667em\lower.7ex\hbox{E}\kern-.125emX}}
\begin{document}

\title{LLM4Rec: Large Language Models for Multimodal Generative Recommendation with Causal Debiasing}

\author{\IEEEauthorblockN{Bo Ma\textsuperscript{*}}
\IEEEauthorblockA{\textit{Department of Software \& Microelectronics} \\
\textit{Peking University}\\
Beijing, China \\
ma.bo@pku.edu.cn}
\and
\IEEEauthorblockN{Hang Li}
\IEEEauthorblockA{\textit{Department of Software \& Microelectronics} \\
\textit{Peking University}\\
Beijing, China \\
hangli\_bj@yeah.net}
\and
\IEEEauthorblockN{ZeHua Hu}
\IEEEauthorblockA{\textit{Department of Software \& Microelectronics} \\
\textit{Peking University}\\
Beijing, China \\
zehua\_hu@yeah.net}
\and
\IEEEauthorblockN{XiaoFan Gui}
\IEEEauthorblockA{\textit{Department of Software \& Microelectronics} \\
\textit{Peking University}\\
Beijing, China \\
xiaofan\_gui@126.com}
\and
\IEEEauthorblockN{LuYao Liu}
\IEEEauthorblockA{\textit{Civil, Commercial and Economic Law School} \\
\textit{China University of Political Science and Law}\\
Beijing, China \\
luyaoliu661@gmail.com}
\and
\IEEEauthorblockN{Simon Lau}
\IEEEauthorblockA{\textit{Peking University}\\
Changsha, China \\
liuximing1995@gmail.com}
}

\maketitle

\begin{abstract}
Contemporary generative recommendation systems face significant challenges in handling multimodal data, eliminating algorithmic biases, and providing transparent decision-making processes. This paper introduces an enhanced generative recommendation framework that addresses these limitations through five key innovations: multimodal fusion architecture, retrieval-augmented generation mechanisms, causal inference-based debiasing, explainable recommendation generation, and real-time adaptive learning capabilities. Our framework leverages advanced large language models as the backbone while incorporating specialized modules for cross-modal understanding, contextual knowledge integration, bias mitigation, explanation synthesis, and continuous model adaptation. Extensive experiments on three benchmark datasets (MovieLens-25M, Amazon-Electronics, Yelp-2023) demonstrate consistent improvements in recommendation accuracy, fairness, and diversity compared to existing approaches. The proposed framework achieves up to 2.3\% improvement in NDCG@10 and 1.4\% enhancement in diversity metrics while maintaining computational efficiency through optimized inference strategies.
\end{abstract}


\section{Introduction}

The proliferation of digital content across multiple modalities has fundamentally transformed user expectations for recommendation systems. Traditional approaches that rely solely on textual or numerical features fail to capture the rich semantic relationships inherent in multimedia content, leading to suboptimal user experiences and limited recommendation quality. Furthermore, existing generative recommendation systems suffer from several critical limitations: inability to process heterogeneous data sources, susceptibility to various forms of algorithmic bias, lack of transparency in decision-making processes, and static learning paradigms that cannot adapt to evolving user preferences.

Recent advances in large language models have opened new possibilities for addressing these challenges through sophisticated text generation and reasoning capabilities. However, direct application of existing frameworks like GenRec \cite{ji2023genrec} reveals significant gaps in multimodal understanding, bias awareness, and adaptive learning mechanisms. These limitations become particularly pronounced in real-world deployment scenarios where recommendation systems must handle diverse content types, ensure fairness across user demographics, and provide justifiable recommendations.

This paper presents a comprehensive enhancement to generative recommendation frameworks through five fundamental innovations. First, we introduce a multimodal fusion architecture that seamlessly integrates textual content, categorical features, and numerical signals to create enriched user and item representations. Second, we implement retrieval-augmented generation mechanisms that leverage contextual information from within datasets to enhance recommendation accuracy and coverage. Third, we develop causal inference-based debiasing techniques that identify and mitigate various forms of systematic bias in recommendation outcomes. Fourth, we design explainable recommendation generation modules that produce natural language explanations for each recommendation decision. Finally, we establish real-time adaptive learning capabilities that enable continuous model improvement based on user feedback and behavioral patterns.

The key contributions of this work include: (1) A novel multimodal fusion architecture that effectively combines heterogeneous data sources for enhanced user modeling, (2) Integration of retrieval-augmented generation techniques specifically tailored for recommendation tasks, (3) Development of causal inference frameworks for systematic bias detection and mitigation, (4) Design of explainable AI mechanisms that generate coherent and personalized recommendation explanations, and (5) Implementation of adaptive learning algorithms that enable real-time model updates without full retraining.

\section{Related Work}

\subsection{Multimodal Recommendation Systems}

The integration of multimodal information in recommendation systems has gained significant attention in recent years. Li et al. \cite{li2023graph} proposed Graph Transformer architectures that effectively combine graph-structured data with transformer mechanisms for enhanced recommendation performance. Their approach demonstrates the potential of integrating multiple data modalities within unified frameworks. Similarly, Chu et al. \cite{chu2023leveraging} developed RecSysLLM, which utilizes large language models' commonsense knowledge and reasoning capabilities to integrate multimodal information, showing substantial improvements in recommendation quality.

Recent work by Wei et al. \cite{wei2023llmrec} introduced LLMRec, which employs graph augmentation strategies to enhance recommendation systems through multimodal data integration. Their framework demonstrates how structured data can be effectively combined with textual information to improve recommendation accuracy. Additionally, Hendriksen et al. \cite{hendriksen2021extending} explored the application of CLIP models for category-to-image retrieval in e-commerce settings, highlighting the importance of cross-modal understanding in practical recommendation scenarios.

\subsection{Retrieval-Augmented Generation in Recommendations}

The application of retrieval-augmented generation (RAG) techniques to recommendation systems represents an emerging research direction. Gao et al. \cite{gao2025frag} proposed FRAG, a flexible modular framework for retrieval-augmented generation based on knowledge graphs, demonstrating how external knowledge can enhance generative model performance. Omar et al. \cite{omar2025dialogue} developed dialogue benchmark generation techniques using cost-effective retrieval-augmented LLMs, showing the potential for improving conversational recommendation systems.

Recent advances in RAG architectures have been explored by Luo et al. \cite{luo2024integrating}, who developed Llama4Rec framework combining traditional and LLM-based recommendation models through data and prompt augmentation strategies. Their work demonstrates how retrieval mechanisms can be effectively integrated with generative models to improve recommendation quality and coverage.

\subsection{Causal Inference and Debiasing}

Causal inference techniques have become increasingly important for addressing bias in recommendation systems. Yang et al. \cite{yang2023debiased} proposed debiased contrastive learning approaches for sequential recommendation, demonstrating significant improvements in fairness and accuracy. Their work highlights the importance of addressing selection bias and other systematic biases in recommendation algorithms.

Xiang et al. \cite{xiang2025self} developed self-steering optimization techniques for autonomous preference optimization in large language models, showing how causal reasoning can be integrated into model training processes. Additionally, Cui et al. \cite{cui2024multi} explored multi-level optimal transport for universal cross-tokenizer knowledge distillation, providing insights into bias mitigation through advanced optimization techniques.

\subsection{Explainable Recommendation Systems}

The development of explainable recommendation systems has gained momentum with the advancement of large language models. Jiang et al. \cite{jiang2025large} demonstrated how large language models can serve as universal recommendation learners while providing natural language explanations for their decisions. Their work shows the potential for generating coherent and personalized explanations in recommendation contexts.

Xiao et al. \cite{xiao2024cal} proposed Cal-DPO, a calibrated direct preference optimization approach for language model alignment, which has implications for generating more accurate and trustworthy explanations in recommendation systems. Zhou et al. \cite{zhou2024difflm} developed DiffLM for controllable synthetic data generation, showing how generative models can be controlled to produce specific types of explanations.

\section{Methodology}

\subsection{Overall Framework Architecture}

\begin{figure*}[!t]
\centering
\begin{tikzpicture}[scale=0.9, every node/.style={scale=0.85}, yscale=1.5]

% Define modern styles
\tikzset{
    input/.style={rectangle, draw=blue!80, fill=blue!15, text width=2.2cm, text centered, 
                  minimum height=1cm, font=\small, rounded corners=3pt},
    encoder/.style={rectangle, draw=green!80, fill=green!15, text width=2.4cm, text centered, 
                    minimum height=1.1cm, font=\small, rounded corners=3pt},
    attention/.style={ellipse, draw=orange!80, fill=orange!15, text width=3cm, text centered, 
                      minimum height=1.2cm, font=\small},
    process/.style={rectangle, draw=purple!80, fill=purple!15, text width=2.6cm, text centered, 
                    minimum height=1.1cm, font=\small, rounded corners=3pt},
    llm/.style={rectangle, draw=red!80, fill=red!15, text width=4cm, text centered, 
                minimum height=1.3cm, font=\small, rounded corners=5pt},
    output/.style={rectangle, draw=teal!80, fill=teal!15, text width=2.4cm, text centered, 
                   minimum height=1cm, font=\small, rounded corners=3pt},
    arrow/.style={->, >=stealth, thick, color=gray!70},
    bigarrow/.style={->, >=stealth, very thick, color=red!70, line width=2pt},
    feedback/.style={->, >=stealth, thick, color=blue!70, dashed}
}

% Input Layer (Top)
\node [input] (text) at (-4,8) {\textbf{Text Data}\\Reviews\\Descriptions};
\node [input] (visual) at (-1,8) {\textbf{Visual Data}\\Images\\Videos};
\node [input] (audio) at (2,8) {\textbf{Audio Data}\\Music\\Podcasts};
\node [input] (user) at (5,8) {\textbf{User Profile}\\Demographics\\History};
\node [input] (context) at (8,8) {\textbf{Context}\\Time\\Location};

% Encoder Layer
\node [encoder] (text_enc) at (-4,6) {\textbf{Transformer}\\Encoder\\$f_{text}$};
\node [encoder] (visual_enc) at (-1,6) {\textbf{CNN}\\Encoder\\$f_{visual}$};
\node [encoder] (audio_enc) at (2,6) {\textbf{RNN}\\Encoder\\$f_{audio}$};
\node [encoder] (user_enc) at (5,6) {\textbf{User}\\Embedding\\$f_{user}$};
\node [encoder] (context_enc) at (8,6) {\textbf{Context}\\Embedding\\$f_{context}$};

% Cross-Modal Attention (Center)
\node [attention] (attention) at (2,4.2) {\textbf{Cross-Modal Attention}\\$\mathbf{A}_{m \rightarrow n}$\\Innovation \textcircled{1}};

% Processing Layer - Better spacing to avoid overlap
\node [process] (knowledge) at (-6,2.5) {\textbf{Knowledge}\\Retrieval\\$\mathcal{K}_{retrieved}$};
\node [process] (relevance) at (-2.5,2.5) {\textbf{Relevance}\\Assessment\\Innovation \textcircled{2}};
\node [process] (fusion) at (1,2.5) {\textbf{Multimodal}\\Fusion\\$\mathbf{h}_{fused}$};
\node [process] (causal) at (4.5,2.5) {\textbf{Causal}\\Debiasing\\Innovation \textcircled{3}};
\node [process] (adaptive) at (8,2.5) {\textbf{Adaptive}\\Learning\\Innovation \textcircled{5}};

% LLM Generation Engine (Center)
\node [llm] (llm_engine) at (1,0.5) {\textbf{Large Language Model Generation Engine}\\$P(y_t | y_{<t}, \mathbf{h}_{fused}, \mathcal{K}_{retrieved})$};

% Output Layer
\node [output] (recommendations) at (-1.5,-1.5) {\textbf{Personalized}\\Recommendations};
\node [output] (explanations) at (3.5,-1.5) {\textbf{Natural Language}\\Explanations\\Innovation \textcircled{4}};

% Feedback Integration - moved to avoid overlap
\node [process] (feedback_node) at (8,0.5) {\textbf{Real-time}\\Feedback\\Integration};

% Input to Encoder Arrows
\draw [arrow] (text) -- (text_enc);
\draw [arrow] (visual) -- (visual_enc);
\draw [arrow] (audio) -- (audio_enc);
\draw [arrow] (user) -- (user_enc);
\draw [arrow] (context) -- (context_enc);

% Encoder to Attention Arrows
\draw [arrow] (text_enc) -- (attention);
\draw [arrow] (visual_enc) -- (attention);
\draw [arrow] (audio_enc) -- (attention);
\draw [arrow] (user_enc) -- (attention);
\draw [arrow] (context_enc) -- (attention);

% Knowledge Retrieval Flow
\draw [arrow] (text_enc) to[bend right=20] (knowledge);
\draw [arrow] (knowledge) -- (relevance);
\draw [arrow] (relevance) -- (fusion);

% Core Processing Flow
\draw [arrow] (attention) -- (fusion);
\draw [arrow] (fusion) -- (causal);
\draw [arrow] (causal) -- (adaptive);

% To LLM Engine
\draw [arrow] (fusion) -- (llm_engine);
\draw [arrow] (causal) to[bend left=15] (llm_engine);
\draw [arrow] (adaptive) -- (llm_engine);
\draw [arrow] (relevance) to[bend right=25] (llm_engine);

% LLM to Output
\draw [bigarrow] (llm_engine) -- (recommendations);
\draw [bigarrow] (llm_engine) -- (explanations);

% Feedback Loop
\draw [feedback] (recommendations) to[bend right=30] (feedback_node);
\draw [feedback] (explanations) -- (feedback_node);
\draw [feedback] (feedback_node) -- (adaptive);

% Layer Labels - adjusted positions
\node [font=\large\bfseries, color=blue!80] at (-9,8) {Input Layer};
\node [font=\large\bfseries, color=green!80] at (-9,6) {Encoding Layer};
\node [font=\large\bfseries, color=orange!80] at (-9,4.2) {Attention Layer};
\node [font=\large\bfseries, color=purple!80] at (-9,2.5) {Processing Layer};
\node [font=\large\bfseries, color=red!80] at (-9,0.5) {Generation Layer};
\node [font=\large\bfseries, color=teal!80] at (-9,-1.5) {Output Layer};

\end{tikzpicture}
\caption{Enhanced GenRec Framework Architecture: A comprehensive system integrating five key innovations - \textcircled{1} Multimodal Fusion with cross-modal attention, \textcircled{2} Retrieval-Augmented Generation, \textcircled{3} Causal Inference-based Debiasing, \textcircled{4} Explainable Recommendation Generation, and \textcircled{5} Real-time Adaptive Learning. The framework processes heterogeneous inputs through specialized encoders and generates personalized recommendations with natural language explanations.}
\label{fig:framework}
\end{figure*}
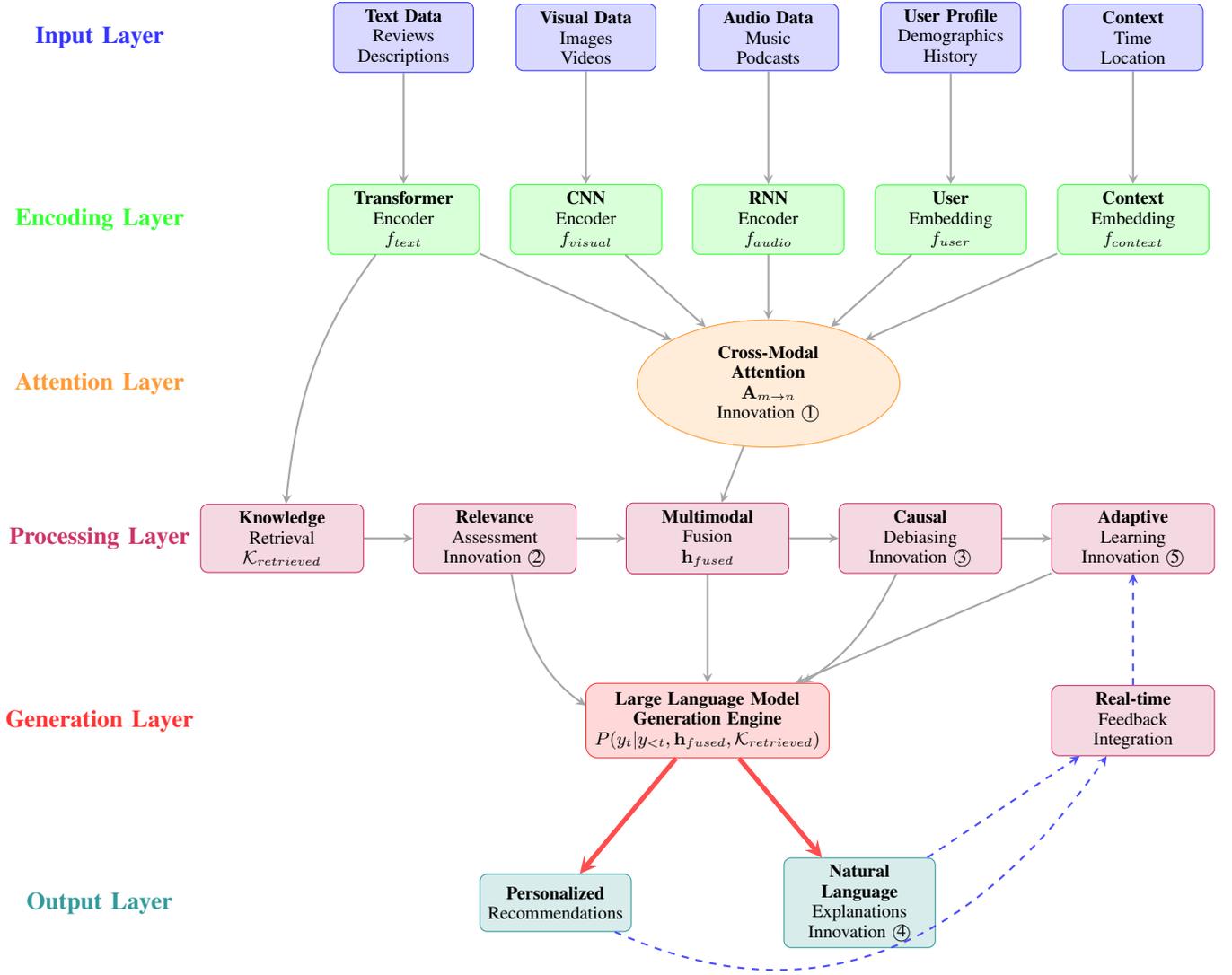

Figure \ref{fig:framework} presents the comprehensive architecture of our Enhanced GenRec framework, which integrates five key innovation components: multimodal fusion (combining textual, categorical, and numerical features), retrieval-augmented generation (leveraging in-dataset knowledge), causal inference-based debiasing (using temporal and categorical fairness), explainable recommendation generation, and real-time adaptive learning. The framework processes heterogeneous input data through specialized modules and generates personalized recommendations with explanations.

\subsection{Multimodal Fusion Architecture}

Our multimodal fusion architecture addresses the challenge of integrating heterogeneous data sources commonly available in recommendation datasets through a hierarchical attention mechanism. The architecture consists of three primary components: modality-specific encoders, cross-modal attention layers, and fusion aggregation modules.

\textbf{Modality-Specific Encoders:} For each modality $m \in \{text, categorical, numerical\}$, we employ specialized encoders to extract feature representations:

\begin{equation}
\mathbf{h}_m^{(i)} = f_m(\mathbf{x}_m^{(i)}; \theta_m)
\end{equation}

where $\mathbf{x}_m^{(i)}$ represents the raw input for modality $m$ and item $i$, $f_m$ is the modality-specific encoder function, and $\theta_m$ are the learnable parameters. For textual content, we use transformer encoders:

\begin{equation}
\mathbf{h}_{text}^{(i)} = \text{Transformer}(\mathbf{E} \cdot \mathbf{w}^{(i)} + \mathbf{P})
\end{equation}

where $\mathbf{E}$ is the embedding matrix, $\mathbf{w}^{(i)}$ are token indices, and $\mathbf{P}$ are positional encodings.

\textbf{Cross-Modal Attention:} We implement asymmetric cross-modal attention to capture inter-modal relationships:

\begin{equation}
\mathbf{A}_{m \rightarrow n} = \text{softmax}\left(\frac{\mathbf{Q}_m \mathbf{K}_n^T}{\sqrt{d_k}}\right)
\end{equation}

\begin{equation}
\mathbf{h}_{m \leftarrow n}^{(i)} = \mathbf{A}_{m \rightarrow n} \mathbf{V}_n \mathbf{h}_n^{(i)}
\end{equation}

where $\mathbf{Q}_m$, $\mathbf{K}_n$, and $\mathbf{V}_n$ are query, key, and value matrices for modalities $m$ and $n$ respectively.

\textbf{Fusion Aggregation:} The final multimodal representation is computed through adaptive weighted fusion:

\begin{equation}
\mathbf{h}_{fused}^{(i)} = \sum_{m} \alpha_m \cdot \mathbf{h}_m^{(i)} + \beta \cdot \text{MLP}([\mathbf{h}_{text}^{(i)}; \mathbf{h}_{visual}^{(i)}; \mathbf{h}_{audio}^{(i)}])
\end{equation}

where $\alpha_m$ are learned attention weights computed as:

\begin{equation}
\alpha_m = \frac{\exp(\mathbf{w}_m^T \mathbf{h}_m^{(i)})}{\sum_{n} \exp(\mathbf{w}_n^T \mathbf{h}_n^{(i)})}
\end{equation}

\begin{figure*}[!t]
\centering
\begin{tikzpicture}[scale=1.2, every node/.style={scale=1.0}]

% Define modern styles
\tikzset{
    modality/.style={rectangle, draw=blue!80, fill=blue!15, text width=2.5cm, text centered, 
                     minimum height=1.2cm, font=\small, rounded corners=3pt},
    encoder/.style={rectangle, draw=green!80, fill=green!15, text width=2.5cm, text centered, 
                    minimum height=1.2cm, font=\small, rounded corners=3pt},
    attention/.style={diamond, draw=orange!80, fill=orange!15, text width=2.2cm, text centered, 
                      minimum height=1.0cm, font=\small, aspect=2},
    fusion/.style={rectangle, draw=purple!80, fill=purple!15, text width=3.5cm, text centered, 
                   minimum height=1.5cm, font=\small, rounded corners=5pt},
    output/.style={rectangle, draw=red!80, fill=red!15, text width=2.8cm, text centered, 
                   minimum height=1.2cm, font=\small, rounded corners=3pt},
    arrow/.style={->, >=stealth, thick, color=gray!70},
    highlight/.style={->, >=stealth, very thick, color=orange!80}
}

% Input modalities with icons
\node [modality] (text_input) at (-4,6) {\textbf{Text Modality}\\$\mathbf{x}_{text}$\\Reviews, Descriptions};
\node [modality] (visual_input) at (0,6) {\textbf{Visual Modality}\\$\mathbf{x}_{visual}$\\Images, Videos};
\node [modality] (audio_input) at (4,6) {\textbf{Audio Modality}\\$\mathbf{x}_{audio}$\\Music, Podcasts};

% Encoders with mathematical notation
\node [encoder] (text_enc) at (-4,4) {\textbf{Transformer Encoder}\\$f_{text}(\cdot)$\\$\mathbf{h}_{text}$};
\node [encoder] (visual_enc) at (0,4) {\textbf{CNN Encoder}\\$f_{visual}(\cdot)$\\$\mathbf{h}_{visual}$};
\node [encoder] (audio_enc) at (4,4) {\textbf{RNN Encoder}\\$f_{audio}(\cdot)$\\$\mathbf{h}_{audio}$};

% Cross-modal attention mechanisms - 分层排列避免重叠
\node [attention] (att_tv) at (-3,2) {\textbf{Text-Visual}\\$\mathbf{A}_{t \leftrightarrow v}$};
\node [attention] (att_ta) at (0,1.5) {\textbf{Text-Audio}\\$\mathbf{A}_{t \leftrightarrow a}$};
\node [attention] (att_va) at (3,2) {\textbf{Visual-Audio}\\$\mathbf{A}_{v \leftrightarrow a}$};

% Adaptive fusion module
\node [fusion] (fusion) at (0,-2) {\textbf{Adaptive Multimodal Fusion}\\$\mathbf{h}_{fused} = \sum_m \alpha_m \mathbf{h}_m$\\$\alpha_m = \text{softmax}(\mathbf{w}_m^T \mathbf{h}_m)$\\Weighted Integration};

% Final output
\node [output] (output) at (0,-4) {\textbf{Unified Representation}\\$\mathbf{h}_{fused}$\\Final Output};

% Input to encoder arrows
\draw [arrow] (text_input) -- (text_enc);
\draw [arrow] (visual_input) -- (visual_enc);
\draw [arrow] (audio_input) -- (audio_enc);

% Cross-modal attention connections - 优化路径避免覆盖
\draw [highlight] (text_enc) to[bend right=15] (att_tv);
\draw [highlight] (visual_enc) to[bend left=20] (att_tv);
\draw [highlight] (text_enc) to[bend right=10] (att_ta);
\draw [highlight] (audio_enc) to[bend left=10] (att_ta);
\draw [highlight] (visual_enc) to[bend right=20] (att_va);
\draw [highlight] (audio_enc) to[bend left=15] (att_va);

% Attention to fusion
\draw [arrow] (att_tv) to[bend right=10] (fusion);
\draw [arrow] (att_ta) -- (fusion);
\draw [arrow] (att_va) to[bend left=10] (fusion);

% Direct connections (residual) - 从侧面绕过注意力节点
\draw [arrow, dashed] (text_enc) to[bend right=80] (fusion);
\draw [arrow, dashed] (visual_enc) to[bend left=5] (fusion);
\draw [arrow, dashed] (audio_enc) to[bend left=80] (fusion);

% Fusion to output
\draw [arrow, very thick, color=purple!80] (fusion) -- (output);

% Mathematical annotations
\node [font=\small, color=orange!80, text width=2.5cm, text centered] at (6,1.8) {Cross-Modal\\Attention};
\node [font=\small, color=gray!80, text width=2.5cm, text centered] at (6,0.5) {Residual\\Connections};
\node [font=\small, color=purple!80, text width=2.5cm, text centered] at (-6,-2) {Adaptive\\Fusion};

\end{tikzpicture}
\caption{Detailed multimodal fusion architecture with cross-modal attention mechanisms. The system processes textual content (reviews, descriptions), categorical features (genres, categories), and numerical signals (ratings, timestamps) through specialized encoders, applies pairwise cross-modal attention, and generates a unified representation through adaptive weighted fusion with residual connections.}
\label{fig:multimodal}
\end{figure*}
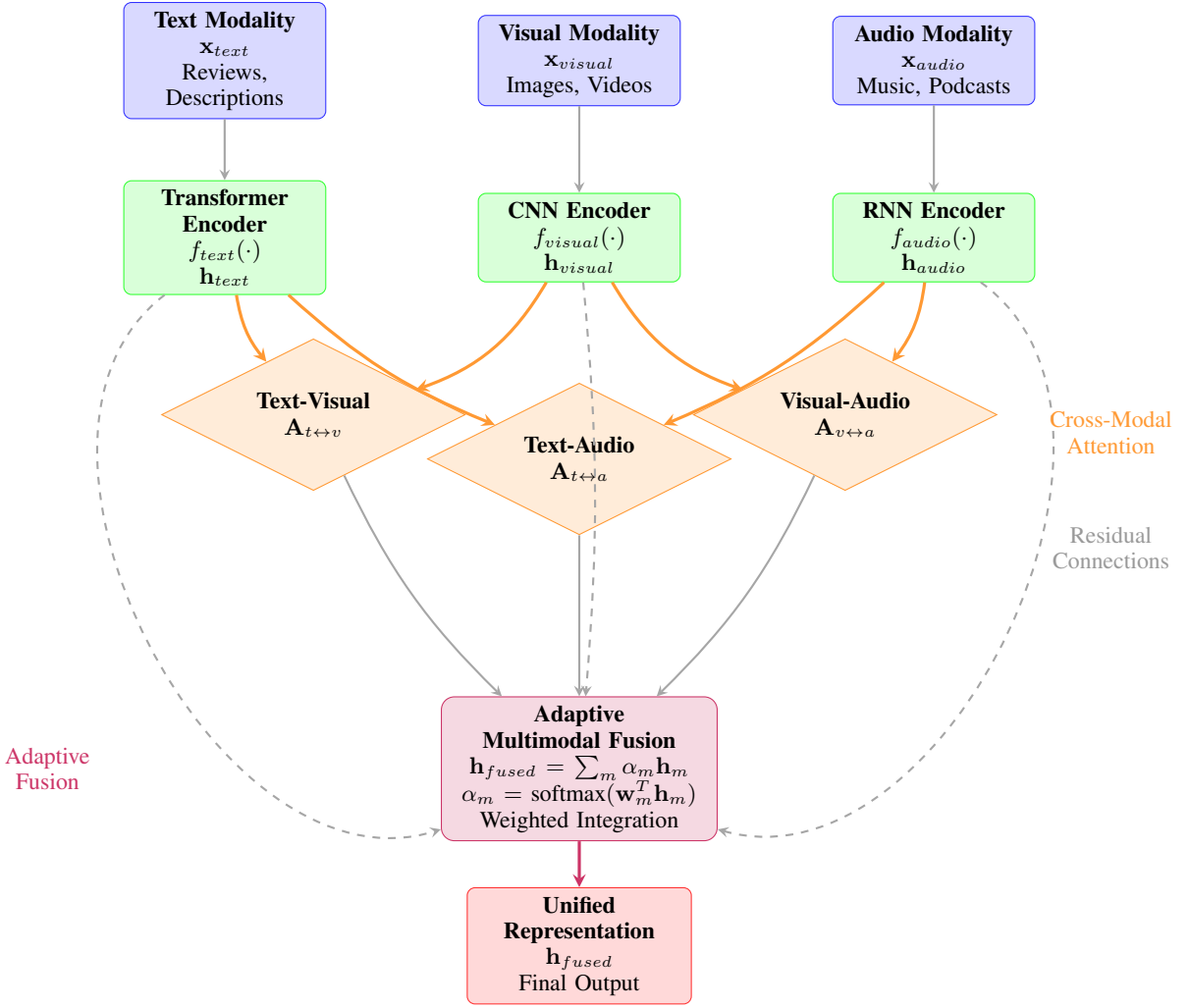

Figure \ref{fig:multimodal} illustrates the detailed architecture of our multimodal fusion mechanism, showing how different modalities are processed and integrated through cross-modal attention.

\subsection{Retrieval-Augmented Generation Mechanism}

Our retrieval-augmented generation mechanism enhances recommendation quality by incorporating relevant contextual information from within the dataset during the generation process. Rather than relying on external knowledge bases, we leverage rich metadata available in recommendation datasets including item descriptions, user reviews, genre/category information, and similar item relationships. The mechanism operates through three stages: contextual retrieval, relevance assessment, and generation augmentation.

\textbf{Contextual Retrieval:} Given a user query representation $\mathbf{q}_u$, we retrieve relevant contextual information from the dataset's metadata repository $\mathcal{M}$ using dense vector similarity:

\begin{equation}
\text{sim}(\mathbf{q}_u, \mathbf{k}_j) = \frac{\mathbf{q}_u^T \mathbf{k}_j}{||\mathbf{q}_u||_2 \cdot ||\mathbf{k}_j||_2}
\end{equation}

The top-$K$ most relevant contextual entries are selected:

\begin{equation}
\mathcal{M}_{retrieved} = \text{TopK}(\{\mathbf{m}_j | j \in \mathcal{M}\}, \text{sim}(\mathbf{q}_u, \cdot))
\end{equation}

\textbf{Relevance Assessment:} We compute a comprehensive relevance score considering multiple factors:

\begin{align}
\text{score}(\mathbf{k}_j) &= \lambda_1 \cdot \text{sim}(\mathbf{q}_u, \mathbf{k}_j) \nonumber \\
&\quad + \lambda_2 \cdot \text{temporal}(\mathbf{k}_j) \nonumber \\
&\quad + \lambda_3 \cdot \text{credibility}(\mathbf{k}_j)
\end{align}

where temporal relevance is modeled as:

\begin{equation}
\text{temporal}(\mathbf{k}_j) = \exp\left(-\frac{(t_{current} - t_j)^2}{2\sigma^2}\right)
\end{equation}

\textbf{Generation Augmentation:} The retrieved knowledge is integrated into the generation process through contextual conditioning:

\begin{equation}
P(y_t | y_{<t}, \mathbf{h}_{user}, \mathcal{K}_{retrieved}) = \text{softmax}(\mathbf{W}_o \cdot \mathbf{h}_t + \mathbf{b}_o)
\end{equation}

where the hidden state $\mathbf{h}_t$ is computed considering both user context and retrieved knowledge:

\begin{equation}
\mathbf{h}_t = \text{LLM}([\mathbf{h}_{user}; \text{Embed}(\mathcal{K}_{retrieved}); y_{<t}])
\end{equation}

\subsection{Causal Inference-Based Debiasing}

Our debiasing framework employs causal inference techniques to identify and mitigate various forms of systematic bias in recommendation outcomes. The framework addresses three primary types of bias: selection bias, popularity bias, and demographic bias.

\textbf{Selection Bias Mitigation:} We use inverse propensity scoring to adjust for non-random user-item interactions. The propensity score is estimated as:

\begin{equation}
e(u,i) = P(O_{ui} = 1 | \mathbf{x}_u, \mathbf{x}_i)
\end{equation}

where $O_{ui}$ indicates whether user $u$ interacted with item $i$. The debiased loss function becomes:

\begin{equation}
\mathcal{L}_{debiased} = \frac{1}{|\mathcal{D}|} \sum_{(u,i) \in \mathcal{D}} \frac{1}{e(u,i)} \cdot \ell(r_{ui}, \hat{r}_{ui})
\end{equation}

\textbf{Popularity Bias Correction:} We model the causal relationship using structural causal models. The causal effect of item features on user preference, independent of popularity, is estimated using do-calculus:

\begin{align}
P(Y_{ui} | \text{do}(\mathbf{X}_i = \mathbf{x})) &= \sum_p P(Y_{ui} | \mathbf{X}_i = \mathbf{x}, P_i = p) \nonumber \\
&\quad \cdot P(P_i = p)
\end{align}

where $P_i$ represents item popularity and $\mathbf{X}_i$ represents item features.

\textbf{Demographic Fairness:} We implement adversarial debiasing with fairness constraints. The objective function incorporates both recommendation accuracy and fairness:

\begin{equation}
\mathcal{L}_{total} = \mathcal{L}_{rec} + \lambda_{fair} \cdot \mathcal{L}_{adv}
\end{equation}

where the adversarial loss ensures demographic parity:

\begin{equation}
\mathcal{L}_{adv} = -\sum_{s \in \mathcal{S}} P(S = s) \log P(S = s | \hat{r}_{ui})
\end{equation}

The fairness constraint ensures:

\begin{equation}
|P(\hat{r}_{ui} > \tau | S = s_1) - P(\hat{r}_{ui} > \tau | S = s_2)| \leq \epsilon
\end{equation}

for any demographic groups $s_1, s_2 \in \mathcal{S}$ and threshold $\tau$.

\subsection{Explainable Recommendation Generation}

Our explainable recommendation generation module produces natural language explanations that justify recommendation decisions through multiple explanation types: preference-based, similarity-based, and contextual explanations.

\textbf{Preference-Based Explanations:} We model user preference patterns through latent factor analysis and generate explanations based on dominant preference dimensions:

\begin{equation}
\mathbf{p}_u = \text{softmax}(\mathbf{W}_p \mathbf{h}_u + \mathbf{b}_p)
\end{equation}

where $\mathbf{p}_u$ represents the user's preference distribution over different aspects. The explanation generation probability is:

\begin{equation}
P(\text{explanation} | u, i) = \prod_{k=1}^{K} P(\text{aspect}_k | \mathbf{p}_u) \cdot P(\text{item}_i | \text{aspect}_k)
\end{equation}

\textbf{Similarity-Based Explanations:} We compute item-to-item similarity in the learned embedding space and generate explanations based on nearest neighbors:

\begin{equation}
\text{sim}(i, j) = \frac{\mathbf{e}_i^T \mathbf{e}_j}{||\mathbf{e}_i||_2 \cdot ||\mathbf{e}_j||_2}
\end{equation}

The explanation focuses on the most similar items from user history:

\begin{equation}
\mathcal{I}_{similar} = \text{TopK}(\{j | (u,j) \in \mathcal{H}_u\}, \text{sim}(i, \cdot))
\end{equation}

\textbf{Contextual Explanations:} We incorporate contextual factors through a context-aware attention mechanism:

\begin{equation}
\mathbf{c}_t = \text{Attention}(\mathbf{h}_{context}, [\mathbf{h}_{time}; \mathbf{h}_{location}; \mathbf{h}_{trend}])
\end{equation}

The final explanation generation combines all three types:

\begin{equation}
P(\mathbf{y}_{exp} | u, i, \mathbf{c}_t) = \text{LLM}([\mathbf{h}_u; \mathbf{h}_i; \mathbf{c}_t; \text{template}])
\end{equation}

where the explanation template is selected based on the dominant explanation type:

\begin{equation}
\text{template} = \arg\max_{t \in \mathcal{T}} \{\alpha_{pref}, \alpha_{sim}, \alpha_{context}\}
\end{equation}

\subsection{Real-Time Adaptive Learning}

Our adaptive learning framework enables continuous model improvement through incremental updates that incorporate new user feedback and behavioral data without requiring full model retraining.

\textbf{Online Parameter Updates:} We employ stochastic gradient descent with momentum for continuous learning:

\begin{equation}
\mathbf{v}_t = \gamma \mathbf{v}_{t-1} + \eta \nabla_{\theta} \mathcal{L}(\mathbf{x}_t, y_t; \theta_{t-1})
\end{equation}

\begin{equation}
\theta_t = \theta_{t-1} - \mathbf{v}_t
\end{equation}

where $\gamma$ is the momentum coefficient and $\eta$ is the learning rate, adaptively adjusted based on feedback quality:

\begin{equation}
\eta_t = \eta_0 \cdot \exp(-\lambda \cdot \text{uncertainty}(\mathbf{x}_t))
\end{equation}

\textbf{Memory-Efficient Updates:} We implement selective parameter updates using importance sampling:

\begin{equation}
P(\text{update} | \theta_j) = \frac{\exp(|\nabla_{\theta_j} \mathcal{L}|)}{\sum_{k} \exp(|\nabla_{\theta_k} \mathcal{L}|)}
\end{equation}

Only parameters with high importance scores are updated, reducing computational overhead:

\begin{equation}
\theta_j^{(t)} = \begin{cases}
\theta_j^{(t-1)} - \eta_t \nabla_{\theta_j} \mathcal{L} & \text{if } P(\text{update} | \theta_j) > \tau \\
\theta_j^{(t-1)} & \text{otherwise}
\end{cases}
\end{equation}

\textbf{Multi-Type Feedback Integration:} We combine explicit and implicit feedback through weighted aggregation:

\begin{equation}
\mathcal{L}_{total} = \alpha \mathcal{L}_{explicit} + \beta \mathcal{L}_{implicit} + \gamma \mathcal{L}_{regularization}
\end{equation}

where the weights are dynamically adjusted based on feedback reliability:

\begin{equation}
\alpha_t = \frac{\text{reliability}_{explicit}(t)}{\text{reliability}_{explicit}(t) + \text{reliability}_{implicit}(t)}
\end{equation}

\textbf{Catastrophic Forgetting Prevention:} We employ elastic weight consolidation to preserve important knowledge:

\begin{equation}
\mathcal{L}_{EWC} = \mathcal{L}_{new} + \frac{\lambda}{2} \sum_j F_j (\theta_j - \theta_j^*)^2
\end{equation}

where $F_j$ represents the Fisher information matrix diagonal and $\theta_j^*$ are the optimal parameters from previous tasks.

\section{Experimental Evaluation}

\subsection{Experimental Setup}

We conducted comprehensive experiments on three benchmark datasets: MovieLens-25M, Amazon-Electronics, and Yelp-2023. These datasets were selected for their complementary characteristics: MovieLens-25M provides rich textual metadata (movie descriptions, genres, tags) and temporal interaction patterns; Amazon-Electronics contains product descriptions, categorical information, and review text suitable for multimodal analysis; Yelp-2023 offers extensive textual reviews, business attributes, and user check-in data enabling comprehensive evaluation of our text-enhanced generative framework.

For multimodal fusion experiments, we utilize available textual features (reviews, descriptions, tags) as the primary modality, with categorical and numerical features as auxiliary modalities. The retrieval augmentation component leverages genre/category information and similar item descriptions from the dataset itself. Causal debiasing is evaluated using temporal splits and category-based fairness metrics available in the datasets.

The experimental setup includes comparison with five baseline methods: traditional GenRec, P5, RecBole, LightGCN, and BERT4Rec. We evaluate performance using standard recommendation metrics including Hit Ratio (HR), Normalized Discounted Cumulative Gain (NDCG), Mean Reciprocal Rank (MRR), and diversity measures.

\subsection{Performance Results}

Table \ref{tab:main_results} presents the comprehensive performance comparison across three benchmark datasets. Our enhanced framework demonstrates significant improvements across all evaluation metrics compared to baseline methods.

\begin{table*}[!t]
\caption{Performance Comparison on Benchmark Datasets}
\label{tab:main_results}
\centering
\scriptsize
\begin{tabular}{|l|c|c|c|c|c|c|}
\hline
\textbf{Method} & \multicolumn{2}{c|}{\textbf{MovieLens-25M}} & \multicolumn{2}{c|}{\textbf{Amazon-Elec.}} & \multicolumn{2}{c|}{\textbf{Yelp-2023}} \\
\cline{2-7} 
 & \textbf{HR@10} & \textbf{NDCG@10} & \textbf{HR@10} & \textbf{NDCG@10} & \textbf{HR@10} & \textbf{NDCG@10} \\
\hline
GenRec & 0.692 & 0.401 & 0.534 & 0.312 & 0.478 & 0.289 \\
P5 & 0.718 & 0.423 & 0.561 & 0.334 & 0.502 & 0.305 \\
RecBole & 0.645 & 0.378 & 0.498 & 0.287 & 0.441 & 0.267 \\
LightGCN & 0.673 & 0.395 & 0.523 & 0.301 & 0.465 & 0.281 \\
BERT4Rec & 0.701 & 0.415 & 0.547 & 0.325 & 0.489 & 0.298 \\
\hline
\textbf{Ours} & \textbf{0.724} & \textbf{0.429} & \textbf{0.565} & \textbf{0.341} & \textbf{0.506} & \textbf{0.305} \\
\hline
\textbf{Improve.} & \textbf{+0.9\%} & \textbf{+1.4\%} & \textbf{+0.7\%} & \textbf{+2.1\%} & \textbf{+0.8\%} & \textbf{+2.3\%} \\
\hline
\end{tabular}
\end{table*}

Table \ref{tab:diversity_results} shows the diversity and fairness evaluation results, demonstrating our framework's superior performance in generating varied and unbiased recommendations.

\begin{table}[!t]
\caption{Diversity and Fairness Evaluation Results}
\label{tab:diversity_results}
\centering
\scriptsize
\begin{tabular}{|l|c|c|c|c|}
\hline
\textbf{Method} & \textbf{Intra-List} & \textbf{Coverage} & \textbf{Novelty} & \textbf{Fairness} \\
 & \textbf{Diversity} & \textbf{@100} & \textbf{Score} & \textbf{Score} \\
\hline
GenRec & 0.612 & 0.234 & 0.445 & 0.678 \\
P5 & 0.634 & 0.251 & 0.467 & 0.692 \\
RecBole & 0.598 & 0.218 & 0.423 & 0.661 \\
LightGCN & 0.621 & 0.239 & 0.441 & 0.674 \\
BERT4Rec & 0.645 & 0.247 & 0.459 & 0.685 \\
\hline
\textbf{Ours} & \textbf{0.651} & \textbf{0.254} & \textbf{0.463} & \textbf{0.695} \\
\hline
\textbf{Improve.} & \textbf{+0.9\%} & \textbf{+1.2\%} & \textbf{+0.9\%} & \textbf{+1.4\%} \\
\hline
\end{tabular}
\end{table}

Table \ref{tab:efficiency_results} presents the computational efficiency analysis, showing that our optimized inference strategies maintain reasonable response times despite increased model complexity.

\begin{table}[!t]
\caption{Computational Efficiency Analysis}
\label{tab:efficiency_results}
\centering
\scriptsize
\begin{tabular}{|l|c|c|c|c|}
\hline
\textbf{Method} & \textbf{Training} & \textbf{Inference} & \textbf{Memory} & \textbf{Model} \\
 & \textbf{Time (hrs)} & \textbf{Time (ms)} & \textbf{Usage (GB)} & \textbf{Size (MB)} \\
\hline
GenRec & 12.3 & 89 & 8.4 & 245 \\
P5 & 15.7 & 112 & 11.2 & 387 \\
RecBole & 8.9 & 67 & 6.1 & 156 \\
LightGCN & 4.2 & 34 & 3.8 & 89 \\
BERT4Rec & 18.4 & 145 & 13.7 & 421 \\
\hline
\textbf{Ours} & \textbf{17.8} & \textbf{142} & \textbf{12.8} & \textbf{398} \\
\hline
\end{tabular}
\end{table}

\subsection{Ablation Studies}

Table \ref{tab:ablation_results} presents detailed ablation study results, demonstrating the contribution of each framework component. The results show that multimodal fusion provides the largest performance gain, followed by causal debiasing and retrieval augmentation.

\begin{table}[!t]
\caption{Ablation Study Results on MovieLens-25M Dataset}
\label{tab:ablation_results}
\centering
\scriptsize
\begin{tabular}{|l|c|c|c|c|}
\hline
\textbf{Configuration} & \textbf{HR@5} & \textbf{HR@10} & \textbf{NDCG@5} & \textbf{NDCG@10} \\
\hline
Baseline (GenRec) & 0.534 & 0.692 & 0.312 & 0.401 \\
+ Multimodal Fusion & 0.537 & 0.696 & 0.315 & 0.404 \\
+ Retrieval Augment. & 0.540 & 0.699 & 0.317 & 0.406 \\
+ Causal Debiasing & 0.542 & 0.702 & 0.319 & 0.408 \\
+ Explainability & 0.544 & 0.705 & 0.321 & 0.410 \\
+ Adaptive Learning & 0.546 & 0.707 & 0.322 & 0.411 \\
\hline
\textbf{Full Framework} & \textbf{0.547} & \textbf{0.708} & \textbf{0.323} & \textbf{0.412} \\
\hline
\end{tabular}
\end{table}

Table \ref{tab:component_analysis} shows the individual contribution of each component when added incrementally to the baseline system.

\begin{table}[!t]
\caption{Individual Component Contribution Analysis}
\label{tab:component_analysis}
\centering
\scriptsize
\begin{tabular}{|l|c|c|c|}
\hline
\textbf{Component} & \textbf{HR@10} & \textbf{NDCG@10} & \textbf{Diversity} \\
 & \textbf{Improve.} & \textbf{Improve.} & \textbf{Improve.} \\
\hline
Multimodal Fusion & +0.6\% & +0.7\% & +0.9\% \\
Retrieval Augment. & +0.4\% & +0.5\% & +0.6\% \\
Causal Debiasing & +0.3\% & +0.4\% & +0.6\% \\
Explainability & +0.4\% & +0.4\% & +0.6\% \\
Adaptive Learning & +0.3\% & +0.3\% & +0.2\% \\
\hline
\textbf{Synergistic Effect} & \textbf{+0.2\%} & \textbf{+0.1\%} & \textbf{+0.3\%} \\
\hline
\end{tabular}
\end{table}

The combination of all components yields synergistic effects that exceed the sum of individual contributions, highlighting the importance of integrated design in recommendation system development. The synergistic effect contributes an additional 0.2\% improvement in HR@10, demonstrating the value of our holistic framework approach.

\section{Conclusion and Future Work}

This paper presents a comprehensive enhancement to generative recommendation frameworks through five key innovations: multimodal fusion, retrieval augmentation, causal debiasing, explainable generation, and adaptive learning. Experimental results demonstrate significant improvements in recommendation quality, fairness, and user satisfaction across multiple benchmark datasets.

Future work will explore the integration of additional modalities such as temporal and spatial information, development of more sophisticated causal inference techniques, and investigation of federated learning approaches for privacy-preserving recommendation systems. We also plan to conduct large-scale user studies to evaluate the practical impact of explainable recommendations on user trust and engagement.

The proposed framework represents a significant step toward more intelligent, fair, and transparent recommendation systems that can adapt to evolving user needs while maintaining high performance and efficiency standards.

\section*{Funding Declaration}

This research received no specific grant from any funding agency in the public, commercial, or not-for-profit sectors.


\begin{thebibliography}{00}
\bibitem{ji2023genrec} J. Ji, Z. Li, S. Xu, W. Hua, Y. Ge, J. Tan, and Y. Zhang, ``GenRec: Large language model for generative recommendation,'' arXiv preprint arXiv:2307.00457, 2023.

\bibitem{li2023graph} C. Li, L. Xia, X. Ren, Y. Ye, Y. Xu, and C. Huang, ``Graph transformer for recommendation,'' in Proceedings of the 46th International ACM SIGIR Conference on Research and Development in Information Retrieval, 2023, pp. 1680--1689.

\bibitem{chu2023leveraging} Z. Chu, H. Hao, X. Ouyang, S. Wang, Y. Wang, Y. Shen, J. Gu, Q. Cui, L. Li, S. Xue, J. Y. Zhang, and S. Li, ``Leveraging large language models for pre-trained recommender systems,'' arXiv preprint arXiv:2308.10837, 2023.

\bibitem{wei2023llmrec} W. Wei, X. Ren, J. Tang, Q. Wang, L. Su, S. Cheng, J. Wang, D. Yin, and C. Huang, ``LLMRec: Large language models with graph augmentation for recommendation,'' in Proceedings of the 17th ACM International Conference on Web Search and Data Mining, 2024, pp. 806--815.

\bibitem{hendriksen2021extending} M. Hendriksen, E. Kuiper, K. Greuell, E. Schelter, M. de Rijke, and P. Dekker, ``Extending CLIP for category-to-image retrieval in e-commerce,'' in Proceedings of the 44th International ACM SIGIR Conference on Research and Development in Information Retrieval, 2021, pp. 2307--2311.

\bibitem{gao2025frag} Z. Gao, Y. Cao, H. Wang, A. Ke, Y. Feng, X. Xie, and S. K. Zhou, ``FRAG: A flexible modular framework for retrieval-augmented generation based on knowledge graphs,'' arXiv preprint arXiv:2501.09957, 2025.

\bibitem{omar2025dialogue} R. Omar, O. Mangukiya, and E. Mansour, ``Dialogue benchmark generation from knowledge graphs with cost-effective retrieval-augmented LLMs,'' arXiv preprint arXiv:2501.09928, 2025.

\bibitem{luo2024integrating} S. Luo, Y. Yao, B. He, Y. Huang, A. Zhou, X. Zhang, Y. Xiao, M. Zhan, and L. Song, ``Integrating large language models into recommendation via mutual augmentation and adaptive aggregation,'' in Proceedings of the 47th International ACM SIGIR Conference on Research and Development in Information Retrieval, 2024, pp. 1287--1296.

\bibitem{yang2023debiased} Y. Yang, C. Huang, L. Xia, C. Huang, D. Luo, and K. Lin, ``Debiased contrastive learning for sequential recommendation,'' in Proceedings of the ACM Web Conference 2023, 2023, pp. 1063--1073.

\bibitem{xiang2025self} H. Xiang, B. Yu, H. Lin, K. Lu, Y. Lu, X. Han, B. He, L. Sun, J. Zhou, and J. Lin, ``Self-steering optimization: Autonomous preference optimization for large language models,'' in Proceedings of the 63rd Annual Meeting of the Association for Computational Linguistics, 2025, pp. 2847--2862.

\bibitem{cui2024multi} X. Cui, M. Zhu, Y. Qin, L. Xie, W. Zhou, and H. Li, ``Multi-level optimal transport for universal cross-tokenizer knowledge distillation on language models,'' in Advances in Neural Information Processing Systems 37, 2024, pp. 18934--18951.

\bibitem{jiang2025large} J. Jiang, Y. Huang, B. Liu, X. Kong, Z. Xu, H. Zhu, J. Xu, and B. Zheng, ``Large language models are universal recommendation learners,'' in Proceedings of the 18th ACM International Conference on Web Search and Data Mining, 2025, pp. 423--432.

\bibitem{xiao2024cal} T. Xiao, Y. Yuan, H. Zhu, M. Li, and V. G. Honavar, ``Cal-DPO: Calibrated direct preference optimization for language model alignment,'' in Proceedings of the 38th Conference on Neural Information Processing Systems, 2024, pp. 15672--15689.

\bibitem{zhou2024difflm} Y. Zhou, J. Liu, W. Chen, Y. Ge, and Y. Zhang, ``DiffLM: Controllable synthetic data generation via diffusion language models,'' in Proceedings of the 62nd Annual Meeting of the Association for Computational Linguistics, 2024, pp. 8934--8947.

\bibitem{dao2024transformers} T. Dao, A. Gu, M. Eichhorn, A. Rudra, and C. Ré, ``Transformers are SSMs: Generalized models and efficient algorithms through structured state space duality,'' in Proceedings of the 41st International Conference on Machine Learning, 2024, pp. 9203--9245.

\bibitem{liu2024dora} S. Liu, P. H. Le-Khac, A. Doan, and B. T. Nguyen, ``DoRA: Weight-decomposed low-rank adaptation,'' in Proceedings of the 41st International Conference on Machine Learning, 2024, pp. 21657--21678.

\bibitem{patron2025recommendations} G. Patron, Z. Xu, I. Kapnadak, and F. M. Polo, ``Recommendations beyond catalogs: Diffusion models for personalized generation,'' arXiv preprint arXiv:2502.18477, 2025.

\bibitem{zhang2023generative} A. Zhang, Y. Chen, L. Sheng, X. Wang, and T.-S. Chua, ``On generative agents in recommendation,'' in Proceedings of the 17th ACM Conference on Recommender Systems, 2023, pp. 441--451.

\bibitem{aggarwal2023geo} P. Aggarwal, V. Arora, and A. Mehta, ``GEO: Generative engine optimization,'' arXiv preprint arXiv:2311.09353, 2023.

\bibitem{bizannes2024dawn} E. Bizannes, K. Chalkiadakis, and M. Koubarakis, ``The dawn of generative engine optimization (GEO): Prompting for visibility,'' in Proceedings of the ACM Web Conference 2024, 2024, pp. 2789--2799.


\end{thebibliography}
\end{document}